# Cross-over behavior of the localized to itinerant transition of 5*f* electrons in the antiferromagnetic Kondo lattice USb$_2$

W. Feng,[1*] D. H. Xie,[1*] X. B. Luo,[1] S. Y. Tan,[1] Y. Liu,[1] Q. Liu,[1] Q. Q. Hao,[1] X. G. Zhu,[1] Q. Zhang,[1] Y. Zhang,[1†] Q. Y. Chen,[1‡] and X. C. Lai[1]

[1]*Science and Technology on Surface Physics and Chemistry Laboratory, Mianyang 621908, China.*

*Correspondence and requests for materials should be addressed to*

Y. Zhang [†] (email: thu_zhangyun@126.com ) and Q.Y.Chen [‡] (email: sheqiuyun@126.com ).

W. Feng[*] and D. H. Xie[*] contributed equally to this work.

In Uranium-based heavy fermion system, the 5*f* electrons display an intermediate character between partial localization and partial itinerancy, which makes the Kondo problem more complicated. Here we use scanning tunneling microscopy/spectroscopy to investigate the (001) surface of the Kondo lattice antiferromagnet USb$_2$. Temperature-dependent *dI/dV* spectra from 4.7 K up to 140 K reveal several peak structures around the Fermi level. Two pronounced peaks are originated from the hybridization between the conduction and 5*f* electrons. We did not observe the mysteriously abrupt change of the electronic state at 45 K, which is previously reported by another STM group and attributed to a novel first-order like transition. Instead, we only observe continuous evolution of the *dI/dV* spectra with temperature. Furthermore, in some scanning regions, we find significant Sb atoms missing from the top layer, which gives us the opportunity to investigate the electronic structure of the U-terminated surface. For the U-terminated surface, an additional sharp peak emerges, which is closely related to the magnetic order.

## I. INTRODUCTION

In intermetallic heavy fermion compounds, containing 4*f* or 5*f* electrons, *f* electrons are essentially localized as magnetic moments at high temperature. As the temperature is lowered, *f* electrons start to hybridize with conduction electrons due to Kondo effect, which tends to form nonmagnetic Fermi-liquid state [1]. Meanwhile, another competing component, namely the Ruderman-Kittel-Kasuya-Yosida (RKKY) interaction, favors a magnetic ordering state [2]. The competition between the Kondo effect and the RKKY interaction determines the ground state, which can be tuned by non-thermal parameters such as pressure, chemical substitution or magnetic field

[3]. In principle, Kondo effect and magnetic ordering compete with each other, and the two channels are both involved with $f$ electrons in heavy fermion systems [4]. Recently, it has been found that the two states can coexist, such as in CeRhIn$_5$ [5, 6] and CeSb [7], more experiments need to be carried out to understand the interplay between Kondo effect and magnetic order and what is the role of $f$ electrons.

USb$_2$ provides an ideal platform to address this complex problem. It crystalizes in the tetragonal structure of anti-Cu$_2$Sb type with a high antiferromagnetic transition temperature ($T_N$) exceeding 200 K [8, 9]. Magnetic moments of U ions are aligned ferromagnetically in the (001) planes, which are stacked along the [001] direction in the antiferromagnetical sequence [10]. de Haas-van Alphen(dHvA) [8, 11] measurements reveal the presence of two-dimensional cylindrical Fermi surface sheets in USb$_2$. By applying pressure, the $T_N$ in USb$_2$ is enhanced first and then an emergent ferromagnetic state appears [12]. Earlier angle-resolved photoemission spectroscopy (ARPES) measurements found a narrow band feature near the Fermi level, which exhibits clear dispersion, indicating that the electronic structure of USb$_2$ has three-dimensional character, although it is a layered compound [13]. Later on kink structure of electron band near the Fermi level was also observed by ARPES measurement in this compound by T. Durakiewicz *et al*. They observed an energy scale of 21 meV and the ultra-small peak width of 3 meV of the kink structure, which may be assigned to interband electron-boson scattering processes [14, 15]. Our previous ARPES results observed two different kinds of nearly flat bands in the antiferromagnetic state of USb$_2$. One is driven by the Kondo interaction between 5$f$ electrons and conduction electrons, and the other originates from the magnetic order [16].

Scanning tunneling microscopy/spectroscopy (STM/STS) is a powerful technique to study both the occupied and unoccupied electronic states of a material, and it has been proven successfully in the study of heavy fermion system, such as imaging of heavy fermions and their formation, heavy-electron superconductivity in both 4$f$- and 5$f$-systems [17-22]. STS measurements of USb$_2$ have been performed by Gianakis *et al.* [23], and they found a novel first order-like transition at 45 K with a drastic change of the spectral feature with temperature, which has not been observed in any of the heavy fermion system before and is rather unexpected.

Here the morphology and electronic structure of the Kondo lattice antiferromagnet USb$_2$ are studied by STM/STS. Two kinds of hybridized quasiparticles, displaying pronounced peak structures, are clearly distinguished. Furthermore, by investigating some regions with significant Sb atoms missing from the top Sb layer, we obtained the electronic feature of the U-terminated surface, where a different sharp peak appears, which is closely related to the magnetic order.

## II. EXPERIMENTAL DETAILS

High quality single crystals of USb$_2$ were grown by the self-flux method. Details of the single-crystal growth, crystal structure and transport measurements can be found in our earlier paper [16]. The single crystals were cleaved in the analyzer chamber with a pressure better than $2\times10^{-11}$ mbar at low temperature, and then immediately transferred to the STM scanning stage. All topographic images were recorded in the constant current mode. The *dI/dV* spectra were collected through a standard lock-in technique by superimposing a small sinusoidal modulation (4 mV, 731 Hz) to the sample bias voltage, then the first-harmonic signal of the current was detected through a lock-in amplifier. Clean tungsten tips were used after e-beam heating and being calibrated on a clean Cu (111) substrate. After each measurement, the tip was rechecked with the Cu (111) surface to exclude possible artifacts from the tip states.

## III. RESULTS AND DISCUSSIONS

Figure 1(a) shows the topographic image of the cleaved surface with a large terrace, and the absence of steps indicates the cleaving process is very successful. Figs. 1(b) and 1(c) present the atomically resolved STM image and line profile, showing square atomic lattice with a spacing of ~0.42 nm, which is consistent with the in-plane lattice constant of the anti-Cu$_2$Sb (P4/nmm) crystal structure of USb$_2$. The homogeneous topography with very small number of defects indicates high sample quality. For USb$_2$, there are several possible surface terminations when the crystal is cleaved. However, in the present study, at least 8 different single crystals were cleaved and we only observed one kind of surface. The earliest density functional theory (DFT) and STM investigation of the surface structure of USb$_2$ has studied the surface energies of various terminations of its (001) surface and revealed that a symmetric cut between Sb planes is the most favorable cleaving plane [24]. In our previous ARPES study of USb$_2$, we also calculated the band structure of USb$_2$ for both U- and Sb-terminations and found that our ARPES results agree well with the band structure for the Sb-terminated surface [16]. In the present study, the cleaving also occurs between the two Sb layers (the green plane in the inset of Fig. 1(a)). The top layer corresponds to the Sb layer (layer 3) exposed upon cleaving, and the second layer is U layer (layer 4) with a spacing of ~0.074 nm between Sb and U layers. Fig. 1(d) shows typical *dI/dV* spectrum measured at 4.7 K, which reveals three pronounced peaks, locating at -20, +6 and +40 meV, respectively. Besides, there is also a hump structure at -60 meV.

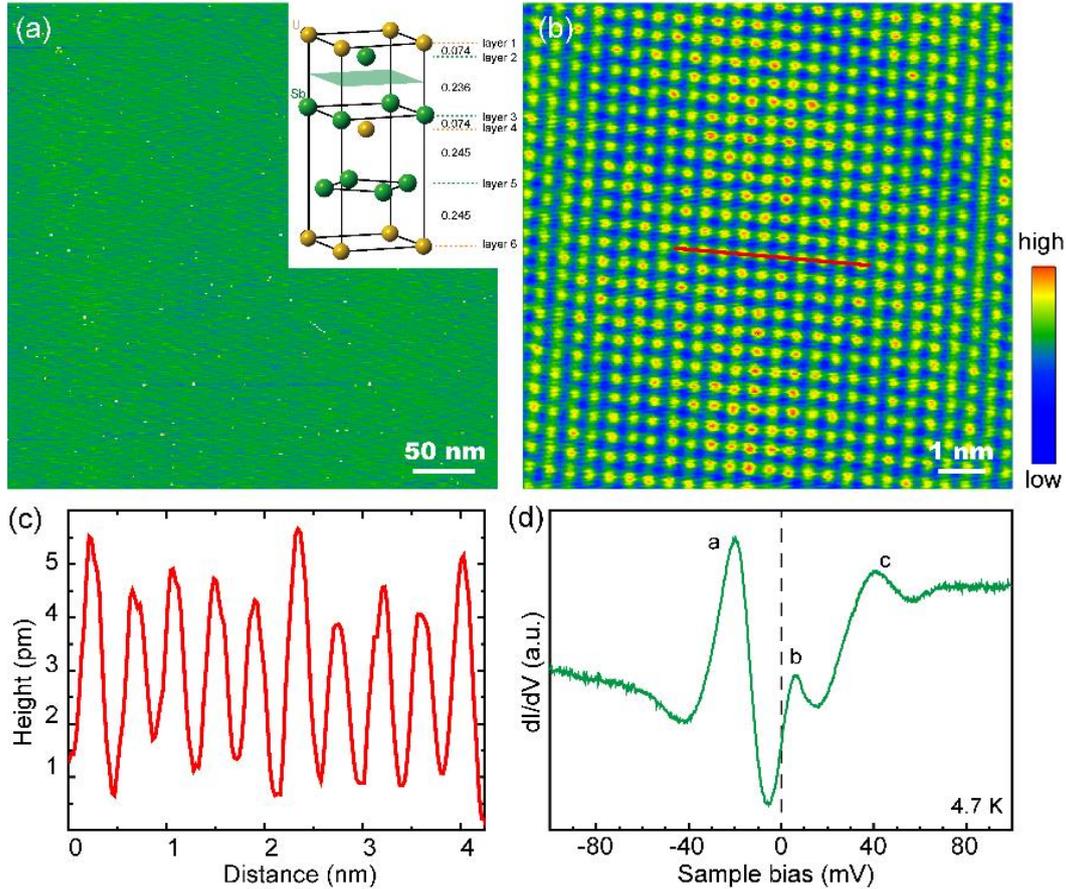

Fig. 1. Topographic images and *dI/dV* spectra of the cleaved surface of $USb_2$. (a) Typical STM constant-current image of the cleaved surface taken at 4.7 K ($V_t$=1.2 V, I=50 pA). The absence of steps indicates high quality and nice cleaving of the samples. The insert shows the layered structure of the crystal $USb_2$. The green plane indicates the cleaving plane in our measurements which is parallel to the (001) surface. The green and yellow dashed lines represent positions of each Sb and U layer. Spacing between adjacent layers is marked with the unit of nm. (b) Atomically resolved STM image of the (001) surface of $USb_2$ ($V_t$=0.2 V, I=200 pA). (c) Height profile measured across the red line in panel (b), which shows the nearest atomic spacing of $USb_2$. (d) Typical *dI/dV* spectrum taken on the perfect (001) surface of $USb_2$ at 4.7 K.

To further reveal the origins of the three peaks, we performed temperature-dependent STS experiments between 4.7 K and 140 K, as shown in Figs. 2(a) and 2(b). When we conduct temperature-dependent STM/STS experiments, the samples were kept for at least two hours for a selected temperature until the temperature no longer changes. In this case, the sample, heater, silicon diode and the STM head all reach thermal equilibrium and have the same temperature. Below the Fermi energy, there is a sharp peak at -20 meV, marked as *a*. With increasing temperature, broadening and weakening of this peak can be clearly observed. At 130 K, peak *a* completely disappears and the spectra show a typical asymmetric lineshape. From our previous ARPES study of USb$_2$, there is a nearly flat η band around the Γ point, locating at 17 meV below the Fermi level. The η band only exists below 130 K and completely disappears at high temperature, which can be ascribed to the Kondo coupling between the *f* band and conduction band at low temperature [16]. The energy position and temperature-dependent behavior of peak *a* observed by STS and the η band revealed by ARPES are nearly the same, indicating that they have the same origin from the *c-f* Kondo entanglement. Compared to peak *a*, peak *b* (at around +6 meV) is not that pronounced and its intensity is relatively weak. The lineshape of peak *b* is also asymmetric as peak *a*, and its energy position is similar to the heavy fermion states observed in other *f*-electron systems, such as URu$_2$Si$_2$ [17, 21] and YbRh$_2$Si$_2$ [20] (just several meV below or above $E_F$), which is attributed to heavy fermion formation. With increasing temperature, peak *b* becomes weak and completely disappears above 77 K. With regard to peak *c*, it displays different behavior as peaks *a* and *b*. It is rather broad and not that sharp as peaks *a* and *b* even at low temperature. With increasing temperature, its energy position stays nearly unchanged and it is still alive at 140 K. The different behavior of peak *c*, compared to peaks *a* and *b*, indicates that it has a different origin and calls for further investigations.

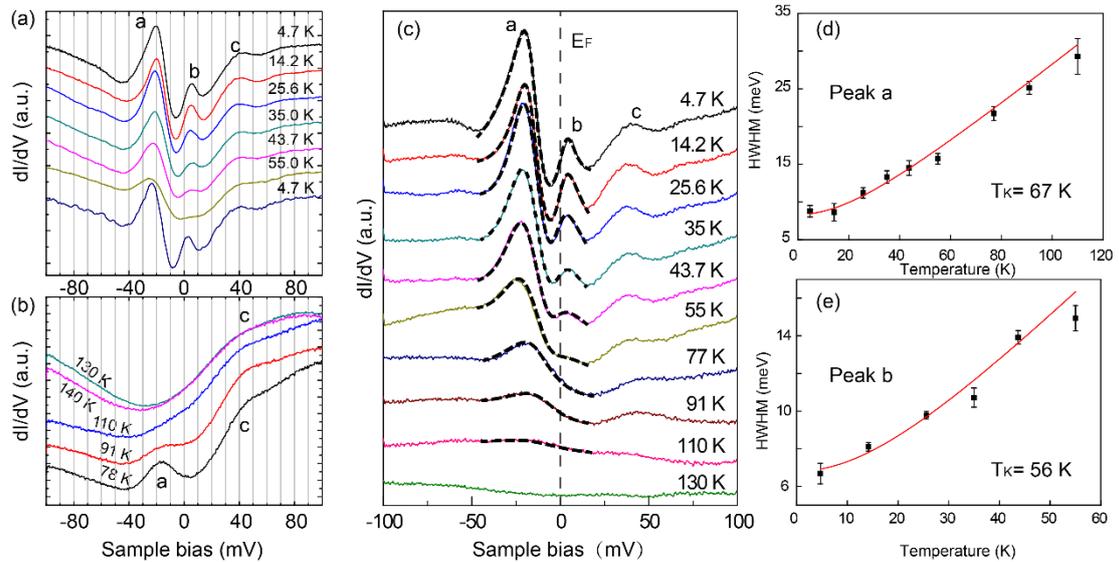

Fig. 2. Temperature-dependent *dI/dV* spectra of the cleaved surface of USb$_2$. (a) Temperature

evolution of the *dI/dV* spectra from 4.7 K to 55 K, and then cooled down to 4.7 K ($V_t$=0.2 V, I=200 pA). (b) Temperature evolution of the *dI/dV* spectra from 77 K to 130 K taken on a different sample ($V_t$=0.1 V, I=300 pA). (c) Processed data of panels (a) and (b) with each spectrum subtracted by the data at 140 K. In order to give an overall impression of the temperature-dependent evolution of the *dI/dV* spectra, data from panels (a) and (b) are put together. The black dashed lines are fits of the spectra by using two Fano line shapes and a linear background. (d, e) Temperature dependence of HWHM of peaks *a* (panel d) and *b* (panel e), respectively. The red curve represents temperature dependence of the width for a single Kondo impurity model.

Peaks *a* and *b* become broad and weak as the temperature increases, and eventually disappear at high temperature. However, the vanishing temperatures of the two peaks seem to be different. At high temperature (above 130 K), the *dI/dV* spectrum presents a broad feature that has weak energy dependence and shows a slightly asymmetric density of states, which is in analogy with that observed in $URu_2Si_2$ [17]. In an effort to quantitatively investigate the evolution of the two peaks with temperature, the spectra obtained at low temperature are subtracted by the spectrum at 140 K, as shown in Fig. 2(c). In order to give an overall impression of the temperature-dependent evolution of the *dI/dV* spectra, data from Figs. 2 (a) and (b) are put together in Fig. 2(c). The line shapes of peaks *a* and *b* are reminiscent of the Fano resonant peaks in the Kondo systems. In a Kondo system, the Fano line shape naturally occurs because of the presence of two interfering tunneling paths from STM tip, one directly into the itinerant electrons, and the other indirectly through the heavy quasiparticles. The Fano resonance has the following function form:

$$dI/dV \propto \frac{(\varepsilon+q)^2}{1+\varepsilon^2}, \quad \varepsilon=\frac{eV-\varepsilon_0}{\Gamma} \qquad (1)$$

Here *q* reflects the quality of the ratio of probabilities between the two tunneling paths, $\varepsilon_0$ is the energy location of the resonance, and $\Gamma$ is the resonance half width at the half maximum (HWHM) [17].

The low-temperature *dI/dV* spectra of $USb_2$ can be well fitted by two Fano lineshape peaks, as shown by the dashed curves in Fig. 2(c). The temperature dependence of HWHMs of the two peaks are displayed in Figs .2 (d) and (e). With increasing temperature, the HWHMs of both peaks increase. Actually results for the single channel spin one-half Kondo impurity model in a Fermi-liquid regime have been used to describe the temperature dependence of HWHM as follows:

$$HWHM = \sqrt{(\pi k_B T)^2 + 2(k_B T_K)^2} \qquad (2)$$

From this equation, the obtained Kondo temperature of peaks *a* and *b* are 67.5 ±1.5 K and 55.9 ±1.7 K, respectively. The success of this model in describing the behavior of both peaks further indicates that the two peaks are related to Kondo physics.

Besides the three peaks in USb$_2$, there is also a hump structure at -60 meV. An earlier study of the crystal field interpretation of UX$_2$ (X=P, As, Sb, Bi) compounds suggests that it is from the crystal field effects [25]. However, it is proposed that this peak is related to the gap opening of the AFM order by STM/STS [23]. In *f*-electron systems, crystal field effects split the *f* states into several energy levels. In USb$_2$, the crystal field tends to split the *f* states into itinerant and localized parts. As for peaks a and b, their temperature evolution and characteristic are related to the Kondo hybridization between *f* electrons and conduction electrons, reflecting the itinerant properties of the *f* electrons. While for the peak at -34 meV observed on the U-terminated layer, it is missing on the Sb-terminated layer and its energy position is consistent with the γ band observed by ARPES, which results from the magnetic order. This state is rather localized. Besides, the peak observed at -60 meV is nearly flat and can still alive above 130 K. It is also a reflection of localized properties of *f* electrons. Since USb$_2$ is less studied, direct observation of the energy levels from crystal field splitting effects are not observed experimentally now by other techniques. It is hard to trace exact contributions to the crystal electric field splittings in USb$_2$. This phenomenon is also observed in SmB$_6$ compound. The crystal field splits the J=5/2 state into $\Gamma_7$, $\Gamma_8(1)$ and $\Gamma_8(2)$ states in SmB$_6$. For $\Gamma_7$ and $\Gamma_8(1)$ states, they are quite itinerant and hybridize with conduction electrons, but $\Gamma_8(2)$ state is quite localized [28, 29].

In the previous STM/STS study of USb$_2$ by a different group [23], they found a novel first-order like electronic transition at 45 K. Such a drastic change of the electronic property with temperature has not been observed in any other heavy-fermion compounds, and is rather mysterious. Rationally, the *f* electrons are completely localized at high temperature. With decreasing temperature, the *f* electrons become itinerant and hybridize with the conduction electrons, forming the heavy quasiparticle band around the Fermi level. This transition of *f* electrons is actually a cross-over behavior and has been verified in many other heavy-fermion compounds [5, 16, 26, 27]. In the current study, in order to exclude possible artificial effects from the STM tips, the observed results were reproduced on at least eight different samples, using five different STM tips. Each STM tip was carefully checked by the surface state of the Cu(111) substrate before performing experiments, and details of the verification of STM tips can be found in Fig. S1 of the Supplemental Material. Temperature-dependent measurements were also reproduced on two other different samples in the liquid helium region and four other different samples in the liquid nitrogen region, which can be found in Figs. S2 and S3 of the Supplemental Material. We did not see that sudden transition at 45 K as they reported. Instead, we only observe continuous evolution of the *dI/dV* spectra with temperature, and this is consistent with the cross-over behavior for the localized-itinerant transition of *f* electrons. We speculate that the sudden change they observed is due to possible artifacts from the STM tip.

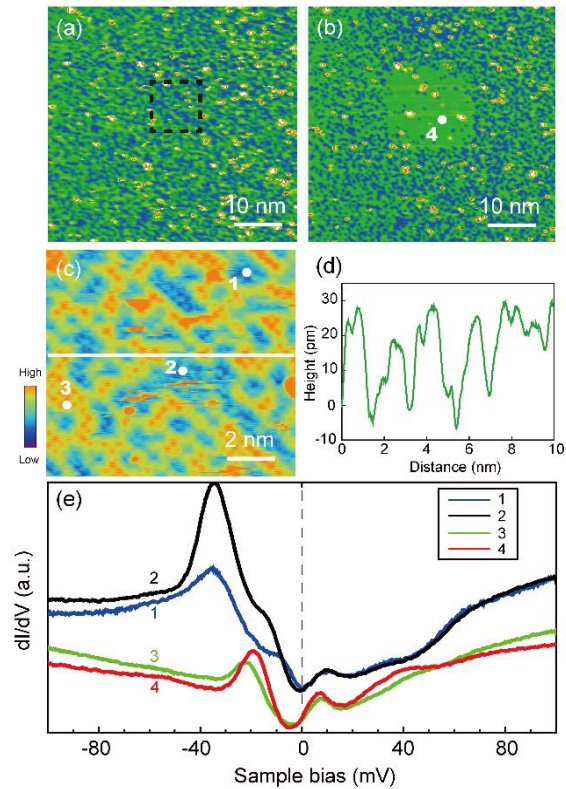

Fig. 3. STM topography and spectroscopy of the cleaved surface with some Sb atoms missing from the top layer. (a) Constant current topographic image of the cleaved surface with a scanning voltage of -0.6 V taken at 4.7 K (I=200 pA). (b) The same region as panel (a) with the same scanning voltage of -0.6 V, but it was taken after the black dashed square region in (a) was scanned with a larger sample voltage of 6 V. After the black dashed square region was scanned with the larger voltage, the residual atoms on the cleaved surface were induced to diffuse and then the ripped surface was repaired as a flat Sb island. (c) Atomic resolution image of the mixed terminated surfaces of USb$_2$ ($V_t$=-0.6 V, I=200 pA). (d) Line profile along the white solid line indicated in (c). (e) *dI/dV* spectra measured at four different positions (denoted as 1 to 4) at 4.7 K, as shown in (b) and (c). Points 1 and 2 are chosen from the regions with more Sb atoms missing from the top layer, and points 3 and 4 are from the Sb atoms and the ripped Sb island, respectively.

As mentioned previously, a symmetric cut between Sb planes is the most favorable cleaving plane, and it makes the study of the electronic states of the U-terminated surface rather difficult. By plenty of trials, we succeeded in finding a region with significant Sb atoms missing from the top Sb layer, which gives us an opportunity to study the spectrum feature of the U-terminated surface. Figs. 3(a-c) show the topographic images of the cleaved surface with a number of atoms missing from the top layer. If the cleaving is between layers 4 and 5 or layers 5 and 6, the exposed two top layers will exhibit a 45 degree rotation of the atomic orientations between the two layers. However, from Fig. 3(c), the square atomic arrangements are clearly visible at both the top and second layers with the same atomic orientation between the two layers. Thus, it can exclude possible cleaving between layers 3 and 4 or layers 4 and 5. Furthermore, the line profile along the white solid line in Fig. 3(c) reveals ~30 pm distance between the top and second layers in Fig. 3(d), which is much smaller than the distance between layers 2 and 3. Consequently, only the cleaving between layers 2 and 3 or layers 5 and 6 could be responsible for the results in Figs. 3(c) and (d), and this is consistent with previous ARPES and DFT calculations [16, 24]. Interestingly, we find that the atoms on the top layer can easily diffuse by changing the scanning voltage. If the scanning voltage is increased from -0.6 to 6 V, the residual atoms on the top layer can gather together and then the ripped surface was repaired as a flat Sb island, which can be traced by the image change from Fig. 3(a) to Fig. 3(b). The *dI/dV* spectrum feature of the repaired Sb flat island is the same as that taken on the Sb-terminated surface, as can be found in Fig. 3(e) (point 4). Since uranium atom is nearly twice heavier than antimony atom, it is more stable than Sb atom and the exposed top layer is more likely to be Sb surface. Consequently, the most favorable cleaving plane is between layers 2 and 3, which is consistent with previous studies [16, 24].

To study the spectrum features of the U-terminated surface, we found some regions where more Sb atoms are missing, such as points 1 and 2 in Figs. 3(b-c), and comparison was made with points 3 and 4 (point 3 is on the Sb atoms, and point 4 is on the repaired Sb island). For the *dI/dV* spectra taken on points 3 and 4, it is almost the same with the spectrum taken on the Sb-terminated surface in Figs.1 and 2. The slight energy shift for point 3 is mainly due to the missing of some adjacent Sb atoms. However, the *dI/dV* spectra exhibit quite different features for points 1 and 2 in Fig.3 (e), where more Sb atoms are missing for the top layer. First, we find a new peak located at about -34 meV. This peak is absent from the Sb-terminated surface, and its strong intensity implies that it may result from the *f* electrons. Its energy position is consistent with the γ band observed by ARPES, which results from the magnetic order [16]. This is natural, since the antiferromagnetic ordering only happens between the U atoms and the antiferromagnetic peak can easily be detected on the U-terminated surface. As for peaks *a* and *b* observed on the Sb-terminated surface, they can also be detected on the U-terminated surface, but their intensity is much lower and also with a small energy shift. This is because the hybridization happens between the *f* electrons and conduction electrons of adjacent Sb and U layers, which makes peaks *a* and *b* also possible to be detected on the U-terminated surface.

From DFT calculations, there are many *f* bands in $USb_2$. ARPES results also reveal that different *f* bands can contribute to various behaviors in $USb_2$. The *f* band around the M point is related to the magnetism, while the one around Γ originates from the *c-f* hybridization [16]. Our STS results provide further details of the electronic structure of $USb_2$ mainly in three aspects: i) The *f* electrons in $USb_2$ show dual characters of itinerancy and localization from STS spectra. ii) The antiferromagnetic order is determined by the U atoms, since the peak related to the antiferromagnetic order is absent on the Sb termination. However, the hybridization of *f* electrons and conduction electrons can happen between adjacent Sb and U layers. iii) We did not observe the mysteriously abrupt change of the electronic state at 45 K. Instead, we only observe continuous evolution of the *dI/dV* spectra with temperature, and this is consistent with the cross-over behavior for the localized-itinerant transition of *f* electrons in heavy fermion compounds.

## IV. CONCLUSION

To summarize, we present a detailed electronic structure study of $USb_2$ by STS. Three peaks can be observed from the spectra at low temperature, and two of which are from the Kondo entanglement between *f* electrons and conduction electrons. Moreover, by investigating some regions with significant Sb atoms missing from the top layer, we obtained the spectrum features for the U- terminated surface, which is closely related to the magnetic order. Our results give a clear microscopic picture of the origins of magnetic order and how magnetic order coexists with the Kondo effect.

## ACKNOWLEDGMENTS

This work was supported by the National Natural Science Foundation of China (No. 12122409, 11974319, 11874330, 11774320, 11904334, 11904335), the National Key Research and Development Program of China (No. 2017YFA0303104) and Science Challenge Project (No. TZ2016004).

[1] R. Jullien, J. Fields and S. Doniach, *Kondo lattice: Real-space renormalization-group approach*, Phys. Rev. Lett. **38**, 1500 (1977).

[2] J. Xavier, R. Pereira, E. Miranda and I. Affleck, *Dimerization induced by the RKKY interaction*, Phys. Rev. Lett. **90**, 247204 (2003).

[3] N. Mott, *Rare-earth compounds with mixed valencies*, Philos. Mag. **30**, 403-416 (1974).

[4] Q. Si, S. Rabello, K. Ingersent and J.L. Smith, *Locally critical quantum phase transitions in strongly correlated metals*, Nature **413**, 804-808 (2001).

[5] Q. Chen, D. Xu, X. Niu, R. Peng, H.C. Xu, C. Wen, X. Liu, L. Shu, S. Tan, X. Lai, Y.Zhang, H.Lee, V.Strocov, F.Bisti, P.Dudin, J. Zhu, H.Yuan, S. Kirchner and D. Feng, *Band dependent interlayer f-electron hybridization in $CeRhIn_5$*, Phys. Rev. Lett. **120**, 066403 (2018).

[6] M. Haze, R. Peters, Y. Torii, T. Suematsu, D. Sano, M. Naritsuka, Y. Kasahara, T. Shibauchi, T. Terashima and Y. Matsuda, *Direct evidence for the existence of heavy quasiparticles in the magnetically ordered phase of $CeRhIn_5$*, J. Phys. Soc. Jpn. **88**, 014706 (2019).

[7] S. Jang, R. Kealhofer, C. John, S. Doyle, J.-S. Hong, J.H. Shim, Q. Si, O. Erten, J.D. Denlinger and J.G. Analytis, *Direct visualization of coexisting channels of interaction in CeSb*, Sci. Adv. **5**, eaat7158 (2019).

[8] D. Aoki, P. Wisniewski, K. Miyake, N. Watanabe, Y. Inada, R. Settai, E. Yamamoto, Y. Haga and Y. Onuki, *Crystal growth and cylindrical Fermi surfaces of $USb_2$*, J. Phys. Soc. Jpn. **68**, 2182-2185 (1999).

[9] R. Wawryk, *Magnetic and transport properties of $UBi_2$ and $USb_2$ single crystals*, Philos. Mag. **86**, 1775-1787 (2006).

[10] M.M. Wysokiński, *Microscopic mechanism for the unusual antiferromagnetic order and the pressure-induced transition to ferromagnetism in $USb_2$*, Phys. Rev. B **97**, 041107 (2018).

[11] D. Aoki, P. Wi´sniewski, K. Miyake, N. Watanabe, Y. Inada, R. Settai, E. Yamamoto, Y. Haga and Y. Onuki, *Cylindrical Fermi surfaces formed by a fiat magnetic Brillouin zone in uranium dipnictides*, Philos. Mag. B **80**, 1517-1544 (2000).

[12] J.R. Jeffries, R.L. Stillwell, S.T. Weir, Y.K. Vohra and N.P. Butch, *Emergent ferromagnetism and T-linear scattering in $USb_2$ at high pressure*, Phys. Rev. B **93**, 184406 (2016).

[13] E. Guziewicz, T. Durakiewicz, M. Butterfield, C. Olson, J. Joyce, A. Arko, J. Sarrao, D. Moore and L. Morales, *Angle-resolved photoemission study of $USb_2$: The 5f band structure*, Phys. Rev. B **69**, 045102 (2004).

[14] T. Durakiewicz, P. Riseborough, C. Olson, J. Joyce, P.M. Oppeneer, S. Elgazzar, E. Bauer, J.

[14] Sarrao, E. Guziewicz and D. Moore, *Observation of a kink in the dispersion of f-electrons*, EPL (Europhysics Letters) **84**, 37003 (2008).

[15] T. Durakiewicz, P. Riseborough and J.-Q. Meng, *Resolving the multi-gap electronic structure of $USb_2$ with interband self-energy*, J. Electron. Spectrosc. **194**, 23-26 (2014).

[16] Q. Chen, X. Luo, D. Xie, M. Li, X. Ji, R. Zhou, Y. Huang, W. Zhang, W. Feng, Y. Zhang, L. Huang, Q. Hao, Q. Liu, X.Zhu, Y, Liu, P.Zhang, X.C.Lai, Q.Si and S. Tan, *Orbital-Selective Kondo Entanglement and Antiferromagnetic Order in $USb_2$*, Phys. Rev. Lett. **123**, 106402 (2019).

[17] P. Aynajian, E.H. da Silva Neto, C.V. Parker, Y. Huang, A. Pasupathy, J. Mydosh and A. Yazdani, *Visualizing the formation of the Kondo lattice and the hidden order in $URu_2Si_2$*, P. Natl. Acad. Sci. USA **107**, 10383-10388 (2010).

[18] A.R. Schmidt, M.H. Hamidian, P. Wahl, F. Meier, A.V. Balatsky, J. Garrett, T.J. Williams, G.M. Luke and J. Davis, *Imaging the Fano lattice to 'hidden order' transition in $URu_2Si_2$*, Nature **465**, 570-576 (2010).

[19] P. Aynajian, E.H. da Silva Neto, A. Gyenis, R.E. Baumbach, J. Thompson, Z. Fisk, E.D. Bauer and A. Yazdani, *Visualizing heavy fermions emerging in a quantum critical Kondo lattice*, Nature **486**, 201-206 (2012).

[20] S. Ernst, S. Kirchner, C. Krellner, C. Geibel, G. Zwicknagl, F. Steglich and S. Wirth, *Emerging local Kondo screening and spatial coherence in the heavy-fermion metal $YbRh_2Si_2$*, Nature **474**, 362-366 (2011).

[21] P. Aynajian, E.H. da Silva Neto, B.B. Zhou, S. Misra, R.E. Baumbach, Z. Fisk, J. Mydosh, J.D. Thompson, E.D. Bauer and A. Yazdani, *Visualizing heavy fermion formation and their unconventional superconductivity in f-electron materials*, J. Phys. Soc. Jpn. **83**, 061008 (2014).

[22] B.B. Zhou, S. Misra, E.H. da Silva Neto, P. Aynajian, R.E. Baumbach, J. Thompson, E.D. Bauer and A. Yazdani, *Visualizing nodal heavy fermion superconductivity in $CeCoIn_5$*, Nat. Phys. **9**, 474-479 (2013).

[23] I. Giannakis, J. Leshen, M. Kavai, S. Ran, C.-J. Kang, S.R. Saha, Y. Zhao, Z. Xu, J. Lynn and L. Miao, *Orbital-selective Kondo lattice and enigmatic f electrons emerging from inside the antiferromagnetic phase of a heavy fermion*, Sci. Adv. **5**, eaaw9061 (2019).

[24] S.-p. Chen, M. Hawley, P. Van Stockum, H. Manoharan and E. Bauer, *Surface structure of cleaved (001) $USb_2$ single crystal*, Philos. Mag. **89**, 1881-1891 (2009).

[25] G. Amoretti, A. Blaise and J. Mulak, *Crystal field interpretation of the magnetic properties of $UX_2$ compounds (X= P, As, Sb, Bi)*, J. Magn. Magn. Mater. **42**, 65-72 (1984).

[26] Q. Chen, D. Xu, X. Niu, J. Jiang, R. Peng, H. Xu, C. Wen, Z. Ding, K. Huang, L. Shu, Y. Zhang, H. Lee, V. Strocov, M. Shi, F. Bisti, T. Schmitt, Y. Huang, P. Dudin, X.C.Lai, S.Kirchner, H.Q.Yuan and D. Feng, *Direct observation of how the heavy-fermion state develops in $CeCoIn_5$*, Phys. Rev. B **96**, 045107 (2017).

[27] R. Zhou, X. Luo, Z. Ding, L. Shu, X. Ji, Z. Zhu, Y. Huang, D. Shen, Z. Liu, Z. Liu, Y. Zhang and Q. Chen. *Electronic structure of $LaIrIn_5$ and f-electron character in its related Ce-115 compounds.* Sci. China-Phys. Mech. Astron. **63**, 117012(2020).

[28] L. Jiao, S. Rößler, D. J. Kim, L. H. Tjeng, Z. Fisk, F. Steglich and S. Wirth. *Additional energy scale in $SmB_6$ at low-temperature*. Nat. Commu.,**7**,13762. (2016).

[29] L. Feng, J. Zhao, H. Weng, Z. Fang and X. Dai. *Correlated topological insulators with mixed valence*. Phys. Rev. Lett. **110**. 096401 (2013).